\documentclass[10pt,twocolumn,letterpaper]{article}

\usepackage[final]{cvpr}      % To produce the REVIEW version
\usepackage{multicol}
\usepackage{multirow}
\usepackage{amsmath}
\usepackage{xcolor}
\usepackage{colortbl}
\definecolor{cvprblue}{rgb}{0.21,0.49,0.74}
\usepackage[pagebackref,breaklinks,colorlinks,allcolors=cvprblue]{hyperref}

\def\paperID{2874} % *** Enter the Paper ID here
\def\confName{CVPR}
\def\confYear{2025}

\title{Integral Fast Fourier Color Constancy}
\author{Wenjun Wei
\and
Yanlin Qian
\and
Huaian Chen
\and
Junkang Dai
\and
Yi Jin
}

\begin{document}
\maketitle

\begin{abstract}
Traditional auto white balance (AWB) algorithms typically assume a single global illuminant source, which leads to color distortions in multi-illuminant scenes. While recent neural network-based methods have shown excellent accuracy in such scenarios, their high parameter count and computational demands limit their practicality for real-time video applications. The Fast Fourier Color Constancy (FFCC) algorithm was proposed for single-illuminant-source scenes, predicting a global illuminant source with high efficiency. However, it cannot be directly applied to multi-illuminant scenarios unless specifically modified. To address this, we propose Integral Fast Fourier Color Constancy (IFFCC), an extension of FFCC tailored for multi-illuminant scenes. IFFCC leverages the proposed integral UV histogram to accelerate histogram computations across all possible regions in Cartesian space and parallelizes Fourier-based convolution operations, resulting in a spatially-smooth illumination map. This approach enables high-accuracy, real-time AWB in multi-illuminant scenes. 
Extensive experiments show that IFFCC achieves accuracy that is on par with or surpasses that of pixel-level neural networks, while reducing the parameter count by over \(400\times\) and processing speed by \(20 - 100\times \) faster than network-based approaches.
\end{abstract}

%传统的自动白平衡（AWB）算法通常假设单一的全局光源，这在多光源场景中会导致颜色失真。虽然近年来基于神经网络的像素级光照估计算法在多光源场景中的准确性表现出色，但其高参数量和计算需求限制了其在实时视频应用中的实用性。快速傅里叶颜色恒常性（FFCC）算法在单一光源场景下被提出，预测单一的全局光源并表现出高效的性能，但其不能直接应用于多光源场景下。为此，我们提出积分快速傅里叶颜色恒常性（IFFCC），IFFCC是FFCC的扩展，专为处理多光源场景而设计。与FFCC不同的是，IFFCC利用所提出的积分 UV 直方图加速了笛卡尔空间中所有可能区域的直方图计算，并行化基于快速傅里叶变换的卷积操作，生成平滑的空间光照图，在多光源场景下实现高精度，实时的自动白平衡。大量实验表明IFFCC的精度可媲美甚至优于像素级神经网络方法，同时将参数量减少了 400 倍以上，速度比基于神经网络的方法快 20–100 倍。

% 传统的自动白平衡（AWB）算法通常假设单一的全局光源，并在整幅图像上应用统一的白平衡增益，这在多光源场景中可能会导致颜色失真。虽然近年来基于神经网络的像素级光照估计算法在准确性方面表现出色，但其高参数量和计算需求限制了其在实时视频应用中的实用性。
% 本研究中，我们提出了一种受快速傅里叶颜色恒常性（FFCC）算法启发的高效解决方案。通过利用积分 UV 直方图，我们的方法加速了笛卡尔空间中所有可能区域的直方图计算，从而实现基于快速傅里叶变换的卷积操作。与 FFCC 不同，我们的方法专为处理多光源场景而设计，生成空间光照图，同时保持精度和实时处理能力。
% 我们将此方法命名为 IFFCC（积分快速傅里叶颜色恒常性），其精度可媲美甚至优于像素级神经网络方法，同时将参数量减少了 400 倍以上，速度比基于神经网络的方法快 20–100 倍。    
\section{Introduction}
% Color constancy is an inherent property of the human vision system that enables perceiving the physical world in a way that remains unaffected by changes in illumination chromaticity. While this ability seems “natural” to the human eye and visual cortex, it presents a significant challenge for computational imaging systems \cite{gijsenij2011computational, foster2011color}. Auto white balance (AWB) in image signal processing (ISP) is essential for estimating ambient illumination chromaticity and ensuring accurate color reproduction.

Color constancy is a fundamental property of the human visual system, allowing for consistent color perception despite changes in illumination chromaticity. While this ability may seem “natural” to the human eye and visual cortex, replicating it in computational imaging systems is a complex challenge \cite{gijsenij2011computational, foster2011color}. Auto white balance (AWB) in image signal processing (ISP) is thus essential for approximating human-like color constancy by estimating ambient illumination chromaticity, enabling accurate color reproduction across varied lighting conditions.

\begin{figure}[t]
    \centering
    \includegraphics[width=\linewidth]{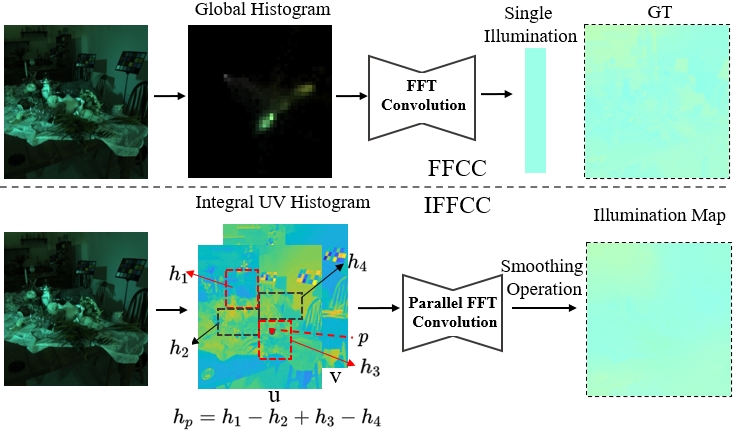}
    \vspace{-5mm}
    \caption{Frameworks comparison between IFFCC and FFCC (top). FFCC uses a global histogram to predict a single illuminant, limiting its application in multi-illuminant scenes. In contrast, IFFCC utilizes an integral UV histogram for efficient region-specific histogram computation, generating spatially continuous illumination map through parallelized Fast Fourier convolution and smoothing to achieve high-accuracy, real-time auto white balance in multi-illuminant scenarios.}
    \label{teaser}
\end{figure}
% Framework comparison between IFFCC and FFCC (top). FFCC uses a global histogram input to predict a single global illuminant. In contrast, IFFCC leverages an integral UV histogram to efficiently compute histograms for any target region using only a few simple operations. Parallelized Fast Fourier convolution then generates multiple localized illumination estimates, followed by a smoothing step to produce a spatially continuous illumination map.

% IFFCC 和 FFCC （上图）框架比较。FFCC 使用全局直方图输入来预测单一的全局光源，不能直接应用于多光源场景除非特定窗口。而 IFFCC 则利用积分 UV 直方图，仅凭几次简单运算即可获取任意目标区域的直方图，同时IFFCC 通过并行化的快速傅里叶卷积生成多个局部光照估计，并通过平滑步骤生成空间连续的光照图，在多光源场景下实现高精度，实时的自动白平衡。

Traditional white balance methods often assume the presence of a single dominant illumination and rely on statistical algorithms \cite{11, cgf, gam, icc}, such as the gray world algorithm \cite{gray} and the white patch algorithm \cite{land1974retinex} etc., to estimate the scene illumination. These algorithms may also be machine learning-based, trained on AWB benchmarks \cite{hu2017fc4, barron2017fast}. However, they frequently bear local color distortion in complex multi-illuminant environments, as a single white balance adjustment is inadequate to handle varying lighting conditions. 
For instance, a room illuminated by both warm high pressure sodium lighting and cool window light may still fail to accurately reproduce the scene's true colors. 
% after global white balance adjustments.

% In recent years, deep neural network-based algorithms have been extensively applied to pixel-level multi-illuminant prediction tasks \cite{cheng2014illuminant, large, cgf, transcc}, leveraging deep learning to accurately estimate illumination map in complex scenes. These methods are particularly effective at capturing edge-aware spatial illumination variations. However, their large model profile and computational demands present significant challenges for deployment in resource-constrained, real-time applications (such as mobile photography devices).
% In a nutshell, more than pixel-wise prediction accuracy, from the industrial perspective, a practical local AWB method better satisfies the following requirements: 

Recently, deep neural network-based algorithms have been widely applied to pixel-level multi-illuminant prediction tasks \cite{cheng2014illuminant, large, cgf, transcc}, utilizing deep learning to achieve accurate illumination map estimation in complex scenes. These methods are especially effective in capturing edge-aware, spatial illumination variations. However, the high parameter counts and computational demands of these models create substantial challenges for deployment in real-time, resource-limited applications, such as mobile photography devices.
Beyond achieving high pixel-level prediction accuracy, a practical local AWB solution from an industrial perspective should prioritize several key requirements:

\noindent \textbf{Effectiveness} - 
% The algorithm can handle various types of light sources and environmental conditions, ensuring effective performance in a wide range of scenarios and including different light intensities and color temperatures.
The algorithm can accomplish the AWB task up to a high accuracy, in any given spatial region, ensuring robust performance in a wide range of scenarios and including different light intensities and color temperatures.

\noindent \textbf{Efficiency} - White balance adjustments must be fast enough to provide immediate feedback during view-finding and capture, particularly in rapidly changing lighting. Limited camera resources require AWB, along with other ISP functions like AE, AF, and color correction, to operate in real time. On resource-constrained devices like smartphones, low computational complexity and memory usage are essential, with AWB processing times ideally under 10 ms.

% \noindent \textbf{Restrained Parameter and Memory} - Low computational complexity and memory usage are necessary for effective performance on resource-constrained devices like smartphones.

\noindent \textbf{Thumbnail Input or Stat} - Many color constancy algorithms require full-resolution, high-bit images, which is resource-intensive on pervasive mobile devices. 
For the purpose of real-time preview, the module for generating AWB stat is often realized in hardware (like DSP), outputting a thumbnail-sized input  (e.g., $64\times48$, 8-bit).  

% \noindent \textbf{Smoothness} - The algorithm should ensure smooth and stable performance under temporally-varying lighting conditions, avoiding artifacts or abrupt transitions. 
% It must be capable of integrating into video streaming, providing smooth and accurate white balance across entire images or video frames.
\noindent \textbf{Smoothness} - The algorithm should maintain stable performance under varying lighting conditions, avoiding artifacts or abrupt transitions, and ensure smooth, accurate white balance across entire images or video frames.

The FFCC algorithm \cite{barron2017fast} effectively meets the requirements for global illumination scenarios but faces challenges in multi-illuminant scenes. Specifically, FFCC requires repeated histogram extraction for each target region, significantly increasing processing time and limiting its suitability for real-time applications, especially on resource-constrained devices like smartphones. Additionally, FFCC lacks mechanisms to ensure smooth spatial transitions under rapidly changing lighting, which can result in abrupt shifts in white balance.

% To overcome FFCC's limitations in multi-illuminant scenarios, we propose Integral Fast Fourier Color Constancy (IFFCC), a localized extension of FFCC \cite{barron2017fast}, specifically designed for white balance in complex lighting environments. Beyond building on FFCC, IFFCC introduces several key contributions:

To overcome FFCC’s limitations in multi-illuminant scenarios, we propose Integral Fast Fourier Color Constancy (IFFCC), a local extension of FFCC \cite{barron2017fast} designed for white balance in complex lighting conditions, as shown in Fig. \ref{teaser}. IFFCC achieves speeds \(20-100 \times\) faster than other local methods, with a runtime of 5.8 ms for \(64 \times 48\) preview images on a CPU. Building upon FFCC, IFFCC introduces several key contributions:

\begin{enumerate}
  \item IFFCC outperforms prior-arts traditional algorithms and most network-based methods in multi-light source scenarios, achieving performance comparable to the best network-based methods;
  % \item We propose the integral UV histogram, which can efficiently compute the histogram for any possible region with just a few simple arithmetic operations, making it \(20-100 \times\) faster than other local methods, with a runtime of 5.8 ms on $64\times48$ preview images using a CPU;
  \item IFFCC leverages the proposed integral UV histogram to accelerate histogram extraction across all possible regions in Cartesian space. It efficiently computes the histogram for any region using only a few simple arithmetic operations, ensuring both high accuracy and real-time performance for multi-illuminant white balance.

  \item IFFCC is spatially adaptive: spatial interpolation and filtering on local estimates ensure smooth transitions while preserving distinct light boundaries and sharp gradients, enabling precise, pixel-level color correction.
\end{enumerate}

  % \item Our method requires only 1.2K parameters (learnable filters and biases), reducing the parameter size by  $>400$ times compared to other methods, significantly easing the computational and memory load on resource-constrained devices;
  % \item It performs interpolation and filtering on multiple predicted illuminants, smoothing illumination transitions between regions while preserving strong light edges or sharp illumination gradients.

% \begin{figure*}
%   \centering
%   \begin{subfigure}{0.68\linewidth}
%     \fbox{\rule{0pt}{2in} \rule{.9\linewidth}{0pt}}
%     \caption{An example of a subfigure.}
%     \label{fig:short-a}
%   \end{subfigure}
%   \hfill
%   \begin{subfigure}{0.28\linewidth}
%     \fbox{\rule{0pt}{2in} \rule{.9\linewidth}{0pt}}
%     \caption{Another example of a subfigure.}
%     \label{fig:short-b}
%   \end{subfigure}
%   \caption{Example of a short caption, which should be centered.}
%   \label{fig:short}
% \end{figure*}

\section{Related Work}
\subsection{Global AWB Methods}

Conventional methods have long served as foundational approaches for uniform color constancy. 
These techniques estimate the illuminant's color based on the statistical characteristics of a single lighting source in the image \cite{9,10,11,gam,icc,14}. 
Notable single-illumination examples include the Gray World assumption \cite{gray}, which posits that the average color of a scene under equal-energy white illumination should appear gray.
Similarly, White Patch \cite{land1974retinex} assumes that the brightest pixel serves as a hint for illumination chromaticity.
Refining these assumptions to focus on local patches or higher-order gradients has led to more robust statistics-based methods, optimized for more-challenging global ambient illuminations. 
These methods include General Gray World \cite{barnard2002comparison}, Gray Edge \cite{14}, Shades-of-Gray \cite{finlayson2004shades}, and LSRS \cite{gao2014efficient}, among others \cite{cheng2014illuminant,Qian_2019_CVPR}.

In same context, more machine learning-based methods have been proposed, such as learning-based kernels \cite{barron2017fast, barron2015convolutional}, convolutional feature extraction \cite{bianco2015color, hu2017fc4}, and other machine learning approaches have been devised specifically for handling single-illumination scenarios \cite{xu2020end,qian2017recurrent,qian2020benchmark,lo2021clcc,liu2019self}. 
The most closely related methods to ours are Barron’s CCC \cite{barron2015convolutional} and FFCC \cite{barron2017fast}. CCC reformulates color constancy as a color localization task in log-chromaticity space, using convolutional kernels to accurately locate the correct color. FFCC builds on this by applying the Fast Fourier Transform on a chromaticity torus, greatly improving the speed and accuracy of chromaticity estimation. However, both are global methods and lack direct applicability to local scenes unless explicitly provided with a local window.

\subsection{Local AWB Methods}
Several multi-illuminant benchmarks have been introduced \cite{gijsenij2011color, large, beigpour2013multi, bianco2014adaptive, bleier2011color, murmann2019dataset, liu2019self}, with various methods proposed to tackle local white balance. These methods often rely on additional prior information or metadata, such as the number of light sources \cite{beigpour2013multi, hsu2008light} or facial regions of interest \cite{bianco2014adaptive}. For instance, \cite{hui2016white} uses flash photography to achieve local white balance. A novel framework in \cite{beigpour2013multi} treats the problem as an energy minimization task within a conditional random field, leveraging local light source estimates. \cite{bleier2011color} employs a global method with spatial variation, while \cite{bianco2017single} introduces a CNN framework based on sliding windows. Additionally, \cite{sidorov2019conditional} applies a generative adversarial network (GAN) to correct images using synthetic datasets, adjusting the color based on the lighting information. More recently, Domislović et al. \cite{domislovic2023color} proposed extracting spatial features for illumination prediction on each patch, and AID \cite{kim2024attentive} introduces a deep model utilizing slot attention \cite{locatello2020object} to generate chromaticity and weight maps for each light source, with the added benefit of allowing user edits.

Among these approaches, most focus on pixel-level predictions, typically using encoder-decoder architectures (e.g., U-Net) to capture fine-grained illumination chromaticity details. However, this approach presents several practical challenges: 
Pixel-level prediction is computationally expensive and can lead to inconsistent results, particularly for high-resolution images, which limits real-time applicability. Additionally, pixel-level color correction may overlook important illumination chromaticity differences that are crucial for accurately rendering color-related edges. As a result, these methods may be less suitable for practical applications or mobile deployment, where computational resources are limited.

% \subsection{Cloest to Ours}
% The cloest methods to ours are Barron's works: CCC \cite{barron2015convolutional} reformulates the color constancy problem as a color localization task in log-chromaticity space, learning convolutional kernels to accurately identify the correct color location. 
% Furthermore, FFCC \cite{barron2017fast}improves computational efficiency by applying the Fast Fourier Transform on the chromaticity torus, significantly enhancing the speed and accuracy of chromaticity estimation.  Till our research, both are global methods and cannot be directly applied locally, unless a local window is offered explicitly.
% Scenarios involving multiple light sources or significant color casts can lead to inaccuracies and color distortions, underscoring the limitations of these single-illumination approaches.

% \subsection{Foundation of FFCC}
\section{Method}
We propose the Integral Fast Fourier Color Constancy (IFFCC) model, which builds upon the foundations of Fast Fourier Color Constancy (FFCC) \cite{barron2017fast}. 
Drawing inspiration from \cite{porikli2005integral}, we implement an histogram search for all possible target regions in image data within Cartesian coordinate space. 
This enables efficient retrieval of any region's histogram in linear computation time, avoiding redundant summation operations. 

\noindent\textbf{FFCC Foundation.}
Consider a linear image stat \( I \) obtained from the camera, where there are no saturated pixels and the black level has been subtracted. 
Assuming the RGB value of any pixel is the product of its true albedo RGB value \( W \) and the uniform illumination value shared across all pixels: $I=W\times L$.
Although this assumption disregards shadows, dichromatic reflections, and spatial variations, it remains effective and widely accepted. Our aim is to estimate the \( L \) given \( I \), and then \(W=I/L\). 
% \begin{eqnarray}
% \forall i \begin{bmatrix}I_{r}^{(i)} 
%  \\I_{g}^{(i)} 
%  \\I_{b}^{(i)} 

% \end{bmatrix} & = & \begin{bmatrix}W_{r}^{(i)} 
%  \\W_{g}^{(i)} 
%  \\W_{b}^{(i)} 

% \end{bmatrix}\circ \begin{bmatrix}L_{r} 
%  \\L_{g}
%  \\L_{b}

% \end{bmatrix}
% \end{eqnarray}
% where \(I^{(i)}\) is the observed color of pixel \(i\), \(W^{(i)}\) is the true white-balanced RGB value of the pixel, and \(L\) is the color of the illuminant.
% Although this assumption disregards shadows, reflections, and spatial variations, it remains an effective and widely accepted model. Our objective is to estimate the value of \( L \) given \( I \), and then \(W=I/L\). 

For an input RGB image \( I \), CCC \cite{barron2015convolutional} defines two log-chroma measures for each pixel \(i\):
\begin{equation}
u^{(i)} = \log(I_{g}^{(i)} / I_{r}^{(i)}) \quad
v^{(i)} = \log(I_{g}^{(i)} / I_{b}^{(i)})
\label{2}
\end{equation}
Neglecting the absolute scaling of \(W\), the estimation of \(L\) is then simplified as $ L_{u}=\log(L_{g}/L_{r})\quad L_{v}=\log(L_{g}/L_{b})$.
% \begin{equation}
%     L_{u}=\log(L_{g}/L_{r}) \quad L_{v}=\log(L_{g}/L_{b})
% \end{equation}
% Our task is to recover \(L_{u}\) and \(L_{v}\). 
Given the unknown absolute scale, the inverse mapping from RGB to UV is challenging. To simplify the recovery process, we assume \(L\) is the unit-norm:
\begin{equation}
\begin{aligned}
    L_{r} = & {\exp(-L_{u})}/{z}  \quad L_{g} = {1}/{z} \quad L_{b} = {\exp(-L_{v})}/{z} \\
    & z = \sqrt{\exp(-L_{u})^{2} + \exp(-L_{v})^{2} + 1}
\end{aligned}
\end{equation}

This logarithmic chromatic space method has several advantages over the RGB approach. First, it involves only two unknowns instead of three, which enhances numerical stability. Additionally, this framework transforms the multiplicative constraints related to \(W\) and \(I\) into additive constraints \cite{finlayson2001color}, effectively converting the color constancy task into a positioning task in a two-dimensional space.

Following FFCC, we use FFT-based convolution in the log-chroma space, leveraging its periodicity to accelerate computations. Given the histogram’s limited size, which does not fully capture the range of colors in natural images, a modular algorithm is needed to allow pixels to "wrap around" in this space:
\begin{equation}
\begin{aligned}
H(x,y) =  & \sum _{i}\left(\mod\left({(u^{(i)} - u_s)}/{b} - x, n\right) < 1 \right. \\
        & \quad \left. \wedge \mod\left({(v^{(i)} - v_s)}/{b} - y, n\right) < 1 \right)
\end{aligned}
\label{6}
\end{equation}
where \(H(x,y)\) represents the count of pixels in \( I \) whose log-chroma coordinates are close to the \( (u, v) \) coordinates associated with the histogram position \( (x, y) \), \(n\) is the number of bins, and \(b\) is the bin size. \( u_s \) and \( v_s \) denote the starting positions of \( u \) and \( v \), respectively. 
The operations here differ from those in CCC. In this case, the \((x, y)\) positions in the histogram no longer represent absolute colors, but rather a set of \((u, v)\) color coordinates. 
The filtered results produce a set of light sources, from which we apply the “gray light de-aliasing” method to resolve the resulting light source clusters. This method assumes that the light sources are as close as possible to the center of the histogram.

FFCC employs the Differentiable Bivariate von Mises (BVM) \cite{mardia1975statistics} method to address the single estimation of light sources from the toroidal PDF. Unlike FFCC, we need to predict multiple light sources for each image. Therefore, the BVM method is parallelized across multiple windows to predict several light sources simultaneously, as described in Section 3.3. The final convolutional structure is as follows:
\begin{equation}
    P=\text{softmax}(B+M\circ \sum_{s}^{}H_s\ast F_s ) 
\end{equation}
where \( B \) is the bias map, \( M \) is the learnable gain map, and \( \{F_s\} \) is the set of learnable filters convolved with the edge-enhanced histogram group \( \{H_s\} \).

\begin{figure*}[t]
    \centering
    \includegraphics[width=0.95\linewidth]{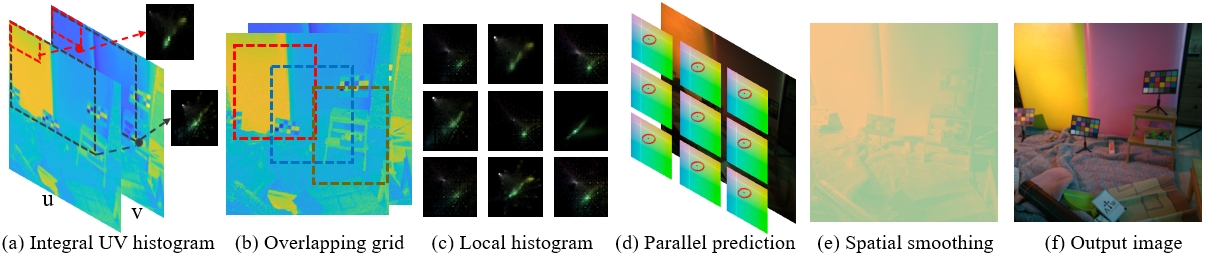}
    \caption{
    Overview of the IFFCC process for predicting multi-illuminant scenes. 
    %To emphasize on the difference compared to FFCC itself, 
    %we mark the different module in bold. 
    We precompute the integral UV histogram of the input image in the log-chroma space (a). Multiple target regions are extracted on the integrated UV histogram with the overlapping grid (b), and the local log-chroma histogram (c) corresponding to each region is computed using simple arithmetic operations on the UV histogram. Each histogram forms a torus that can be unfolded, producing a cluster of potential light sources. A de-aliasing step selects the most likely light source for each region (d), represented by a point within an ellipse, where the ellipse indicates IFFCC’s covariance output. Final light estimates are smoothed through interpolation and guided filtering, creating a continuous illumination map (e) that combines with the input to produce a white-balanced image (f) for multi-illuminant scenes.}
    \label{bone}
\end{figure*}

\subsection{Integral UV Histogram}
FFCC predicts a global illumination for the image. However, when multiple illuminations are anticipated for different target regions, the process requires FFCC to repeatedly extract histograms from these various areas, significantly increasing computational load and processing time. Inspired by the work of \cite{porikli2005integral}, we efficiently compute the histograms for all potential regions within a given image using Cartesian coordinates. By leveraging the spatial positions of data in this coordinate system, we propagate a function starting from the origin to navigate through the remaining positions. The histogram for the current point is then updated based on the histograms of previously processed points. Once the integral histogram for each data point is obtained, the histogram for the target region can be derived using simple addition and subtraction operations. This method eliminates the need to reconstruct separate histograms for each region individually. As a result, this efficient approach significantly reduces computational overhead and improves processing time, enabling rapid analysis of multiple areas within the image. 

After transforming the image into log-chroma space by Eq.\ref{2}, we define the integral uv histogram as follows:
\begin{equation}
   H_{\text{Integral}}(x^{n}, b) = \bigcup_{m=0}^{n} B(u, v)^{m} 
\end{equation}
\begin{equation}
   B(u, v) = \left( 
   \begin{aligned}
   &\mod\left( \text{round}\left(\frac{u - u_{s}}{b}\right), n \right), \\
   & \mod\left( \text{round}\left(\frac{v - v_{s}}{b}\right), n \right)
   \end{aligned}
   \right)
\end{equation}
where \( x^n \) is a sequence of points \(x^0, x^1, \dots, x^n \). \( B(.) \) represents the corresponding bin for the current data point, \( \bigcup \) denotes the union operator, and the value of bin \( b \) in \( H_{\text{Integral}}(x^n, b) \) is the sum of the histogram values of bin \( b \) from all previously traversed data points. 
The integral histogram can also be expressed recursively as follows:
\begin{equation}
    H_{\text{Integral}}(x^{m},b)= H_{\text{Integral}}(x^{m-1},b)\bigcup_{}^{}B(u,v)^{m}
\end{equation}
where at the initial data point, \( H_{\text{Integral}}(0, b) = 0 \).

In log-chroma space, updating the histogram at the current point requires the histograms of the data points above, the left, and the top-left of the current point. These three histograms have already been calculated in previous scans, so only three simple arithmetic operations are needed to update the histogram for the current point. Afterward, the bin value \( b \) for the current point is calculated, and the histogram is updated accordingly. Before each propagation step, all scanned bin values are copied to the current bin values, which can be efficiently handled through fast hardware or pointer operations. 

For an RGB image converted to \( N_1 \times N_2 \times 2 \) log-chroma space data, the integral histogram propagated via wavefront scanning from the top-left starting point can be written as:
\begin{equation}
\begin{aligned}
H_{\text{Integral}}(u, & v, b) = H_{\text{Integral}}(u - 1, v, b) \\
& + H_{\text{Integral}}(u, v - 1, b) \\
& - H_{\text{Integral}}(u - 1, v - 1, b) + B(u, v)
\end{aligned}
\label{15}
\end{equation}
where \((u,v)\) is the coordinate of the log-chroma space. In our method, the data in log-chroma space can be represented by a two-dimensional floating point tensor of a two-dimensional scene \cite{porikli2005integral}, allowing us to calculate the computational load ratio between the traditional histogram method and the integral histogram method as follows:
\begin{equation}
    R=\frac{[217M_1M_2+4B^2]S_1S_2}{274+11(68+4B^2)S_1S_2} 
\end{equation}
where \( M_1 \) and \( M_2 \) represent the size of the target region in the two-dimensional scene, \( B \) is the number of bins, and \( S_1 \) and \( S_2 \) are the different scales used when extracting histograms. The result of the calculation for \( R \) when \( M_1 = M_2 = 128 \), \( B = 64 \), and \( S_1 = S_2 = 1 \) is approximately \( R = 19.71 \).

% \subsection{Sliding Window with Overlap}
% \subsection{Localizing Inference}
\subsection{Overlapping Grid}  
During inference, we use a sliding window with overlap to efficiently retrieve histograms for sub-regions in the log-chroma space, as illustrated in Fig.\ref{bone}. By pre-computing an integral histogram over the entire scene, histograms for any arbitrary sub-region can be directly retrieved using Eq.\ref{15}, without reconstructing each one from scratch, which significantly reduces computational costs. This overlapping grid method captures local chromatic variations by sliding a defined window across the scene and calculating histograms incrementally. The overlapping regions between windows maintain continuity and minimize edge effects, enhancing the accuracy of chromatic analysis throughout the scene.

% \subsection{One Inference, Multi illumination}
\subsection{Parallel Prediction}
To speed up inference, our method requires only one inference to predict several illumination across multiple target regions within the image. Given an image's integral UV histogram in log-chroma space, denoted as \( H_{\text{integral}}(H, W, B, B, N) \), where \( H \) and \( W \) represent the image size, \( B \) is the number of bins, and \(N\) is the number of augmented channel’s histogram. The overlapping grid approach can be applied to obtain a collection \( X \) of \( k \) target region histograms. This collection \( X \), represented as \( X(k, B, B, N) \), includes histograms for each specific target region in the log-chroma space. The process of calculating the PDF of \( X \) in the frequency domain is as follows:
\begin{equation}
    H = \text{IFFT}\left(\sum_{k}\text{FFT}(X) \cdot \text{FFT}(F)\right) + B
\end{equation}
\begin{equation}
P = \frac{\text{exp} ^{(H - \max(H))}}{\sum_{\text{dims}} \text{exp}^{(H - \max(H))}}
\end{equation}
where \(\text{FFT}(.)\) and \(\text{IFFT}(.)\) are the fast Fourier transform and its inverse, respectively, \( F \) is the convolution kernel, and \( B \) is the bias map. 

To parallelize the estimation of the mean of the Bivariate Von Mises (BVM) distribution from a histogram, we calculate circular mean values for the histogram coordinates \( i \) and \( j \) across multiple regions concurrently. This process enables effective mean direction estimation within the toroidal data space, essential for capturing periodic patterns in wrapped grid data. For each region \( k \), we define \( \mu = \{ \mu_k \}_{k=1}^{K} \), where each \( \mu_k \) is given by:
\begin{equation}
    \mu_k = \begin{bmatrix} u_{s} \\ v_{s} \end{bmatrix} + b \begin{bmatrix} \text{mod}\left(\frac{n}{2\pi} \arctan2( y_{1}^{k},  x_{1}^{k}), n\right) \\ \text{mod}\left(\frac{n}{2\pi} \arctan2( y_{2}^{k},  x_{2}^{k}), n\right) \end{bmatrix}
\end{equation}
where the components \( y_{1}^{k} \), \( x_{1}^{k} \), \( y_{2}^{k} \), and \( x_{2}^{k} \) are computed in parallel for each \( k \):
\begin{equation}
\begin{aligned}
    y_{1}^{k} &= \sum_{i} P^{k}_{i} \sin\left(\frac{2\pi i}{n}\right), x_{1}^{k} = \sum_{i} P^{k}_{i} \cos\left(\frac{2\pi i}{n}\right) \\
    y_{2}^{k} &= \sum_{j} P^{k}_{j} \sin\left(\frac{2\pi j}{n}\right), x_{2}^{k} = \sum_{j} P^{k}_{j} \cos\left(\frac{2\pi j}{n}\right)
\end{aligned}
\end{equation}
where \( P^{k}_{i} \) and \( P^{k}_{j} \) represent the marginal distributions for rows and columns of each target region's 2D histogram. After concurrently obtaining illumination values for all \( k \) target regions, we reconstruct the full illumination map by interpolating or applying edge-preserving filtering methods.

% To estimate the mean of the Bivariate Von Mises (BVM) distribution from a histogram, one can calculate the circular mean values for the histogram coordinates \( i \) and \( j \). By determining these values on the circular scale, we obtain an effective estimate for the mean direction within the toroidal data space, which is essential for accurately capturing periodic patterns in wrapped grid data. We can define \( \mu = \{ \mu_k \}_{k=1}^{K} \). Thus, \( \mu \) is the set \( \{ \mu_1, \mu_2, \ldots, \mu_K \} \), where each \( \mu_k \) is given by:
% \begin{equation}
%     \mu_k = \begin{bmatrix} u_{s} \\ v_{s} \end{bmatrix} + b \begin{bmatrix} \text{mod}\left(\frac{n}{2\pi} \arctan2( y_{1}^{k},  x_{1}^{k}), n\right) \\ \text{mod}\left(\frac{n}{2\pi} \arctan2( y_{2}^{k},  x_{2}^{k}), n\right) \end{bmatrix}
% \end{equation}
% \begin{equation}
% \begin{aligned}
%     y_{1}^{k} &= \sum_{i} P^{k}_{i} \sin\left(\frac{2\pi i}{n}\right), x_{1}^{k} = \sum_{i} P^{k}_{i} \cos\left(\frac{2\pi i}{n}\right) \\
%     y_{2}^{k} &= \sum_{j} P^{k}_{j} \sin\left(\frac{2\pi j}{n}\right), x_{2}^{k} = \sum_{j} P^{k}_{j} \cos\left(\frac{2\pi j}{n}\right)
% \end{aligned}
% \end{equation}
% where \( P^{k}_{i} \) and \( P^{k}_{j} \) represent the marginal distributions of the 2D histogram for rows and columns, respectively. After obtaining the illumination values for \( k \) target regions, we can recover the illumination map for the entire image using interpolation or some edge-preserving filtering methods.

% \subsection{Spatial Smoothing (interpolation or edge-preserving filter)}
\subsection{Spatial Smoothing}
In multi-light scenarios, patch-wise illumination estimates often exhibit discrete, locally discontinuous characteristics. To reconstruct a smooth and seamless illumination map across the entire image, it is vital to postprocess these patch-level estimates using interpolation. 
% Interpolation methods can effectively fill the gaps between patches, providing smooth transitions that connect each local region seamlessly. 

% Interpolation methods differ in complexity and smoothness. Linear interpolation offers a straightforward approach by connecting known points with straight lines, creating a piecewise linear transition for estimating intermediate values. In contrast, polynomial interpolation employs higher-degree polynomials for smoother transitions, but excessively high degrees can introduce oscillations that compromise accuracy. After obtaining illumination predictions for \( k \) target regions, interpolation adjusts these values based on their relative positions, resulting in a smooth illumination map. However, while interpolation enhances overall smoothness, it may also blur local details, particularly in areas with strong light edges or sharp illumination gradients, potentially losing some edge information.

We employ linear interpolation due to its simplicity and ability to create smooth, piecewise-linear transitions for estimating intermediate values. Although higher-degree polynomial interpolation could be used, it risks introducing oscillations. 
After getting estimates for \( k \) target regions, interpolation adjusts these values based on their relative locations, resulting in a smooth illumination map. 
Although interpolation enhances overall smoothness, it may blur local details, particularly in regions with strong light edges or steep illumination gradients, potentially leading to a loss of edge information.
% Till now , overall smoothness is obtained, but with a risk of losing local details.

Guided filtering \cite{he2012guided} is then applied to preserve details at the boundaries and structural features of the illumination map, preventing blur or distortion during transitions. Overall, combining these techniques results in a globally coherent illumination distribution that maintains edge clarity while appearing smoother and more natural. Given the interpolated illumination map \( L \) and the raw image \( G \) as the guiding image, the output illumination map \( O \) at each pixel \( O_i \) can be expressed as:
\begin{equation}
O_i = a_k G_i + b_k, \quad \forall i \in \omega_k
\end{equation}
\begin{equation}
a_k = \frac{\frac{1}{|\omega|} \sum_{i \in \omega_k} G_i L_i - \bar{G_k} \bar{L_k}} {\sigma_k^2 + \epsilon}, \quad
b_k = \bar{L_k} - a_k \bar{G_k}
\end{equation}
here, \( \bar{G_k} \) and \( \bar{L_k} \) are the mean values of \( G \) and \( L \) in window \( \omega_k \), with \( \sigma_k^2 \) as the variance of \( G \). The constant \( \epsilon \) prevents division by zero. The final illumination map \( O \) is obtained by weighted averaging.
\begin{equation}
O = \frac{1}{|\omega|} \sum_{k \in \omega_i} (a_k G_i + b_k)
\end{equation}

% By employing these methods, we can effectively create a coherent illumination map that accurately reflects the lighting conditions across the entire image while preserving critical features. 
% This post-processing step is essential for enhancing the overall visual quality and ensuring that the reconstructed illumination aligns well with the underlying content of the image.
This post-processing step is essential: it creates a coherent illumination map that accurately reflects the lighting conditions across the entire image while preserving critical features, thus the reconstructed illumination aligns well with the underlying content of the image.

\begin{table*}[ht]
\centering
\resizebox{0.95\linewidth}{!}{
\setlength{\tabcolsep}{3.2pt}
\renewcommand{\arraystretch}{0.9}
\begin{tabular}{lcccccccccccc}
\hline
\multicolumn{1}{c}{\multirow{2}{*}{Method}}                            & \multicolumn{2}{c}{Canon\_5d} & \multicolumn{2}{c}{Canon\_550d} & \multicolumn{2}{c}{Moto}      & \multicolumn{2}{c}{Panasonic} & \multicolumn{2}{c}{Sony}      & \multicolumn{2}{c}{Use all}   \\ \cline{2-13} 
\multicolumn{1}{c}{}                                                   & mean          & median        & mean           & median         & mean          & median        & mean          & median        & mean          & median        & mean          & median        \\ \hline
Huassin and Akbari \cite{hussain2018color}                                                & 13.88         & 13.76         & 13.73          & 13.54          & 13.06         & 12.89         & 13.62         & 13.73         & 14.22         & 13.68         & 13.67         & 13.46         \\
CRF(White-Patch)\cite{beigpour2013multi}                                                       & 7.66          & 5.96          & 7.69           & 5.77           & 6.94          & 5.32          & 7.58          & 5.69          & 7.90          & 5.86          & 7.19          & 5.44          \\
Patch-based\cite{gijsenij2011color}                                                 & 4.85          & 3.11          & 4.61           & 2.98           & 4.17          & 2.49          & 4.58          & 2.95          & 4.96          & 3.16          & 4.30          & 2.89          \\
\begin{tabular}[c]{@{}l@{}}Keypoint-based\cite{gijsenij2011color}\end{tabular} & 5.59          & 3.88          & 5.48           & 3.94           & 4.97          & 3.28          & 5.42          & 3.79          & 5.76          & 3.82          & 5.46          & 3.59          \\
Superpixel-based\cite{gijsenij2011color}                                                      & 4.88          & 3.68          & 4.79           & 3.56           & 4.19          & 2.89          & 4.55          & 3.54          & 4.97          & 3.82          & 4.20          & 3.10          \\
IFFCC                                                                  & \cellcolor{pink!60}{\textbf{2.06}} & \cellcolor{pink!60}{\textbf{1.54}} & \cellcolor{pink!60}{\textbf{2.16}}  & \cellcolor{pink!60}{\textbf{1.77}}  & \cellcolor{pink!60}{\textbf{1.90}} & \cellcolor{pink!60}{\textbf{1.36}} & \cellcolor{pink!60}\textbf{{2.06}} & \cellcolor{pink!60}{\textbf{1.59}} & \cellcolor{pink!60}{\textbf{2.66}} & \cellcolor{pink!60}{\textbf{1.96}} & \cellcolor{pink!60}{\textbf{2.19}}          & \cellcolor{pink!60}{\textbf{1.56}}\\ \hline
Bianco et al.\cite{bianco2017single}                                                          & 4.63          & 4.02          & 4.59           & 4.16           & 4.27          & 3.89          & 4.98          & 4.56          & 5.34          & 4.76          & 8.01          & 5.69          \\
HypNet/SelNet\cite{shi2016deep}                                                          & 5.42          & 4.93          & 5.16           & 4.83           & 5.09          & 4.74          & 5.23          & 4.98          & 5.82          & 5.34          & 6.31          & 3.95          \\
Domislovic et al.\cite{domislovic2021outdoor}                                                      & \cellcolor{yellow!60}{2.63}          & \cellcolor{yellow!60}{2.18}          & \cellcolor{yellow!60}{2.77}           & \cellcolor{yellow!60}{2.06}           & \cellcolor{yellow!60}{2.07}          & \cellcolor{yellow!60}{1.49}          & \cellcolor{yellow!60}{2.38}          & \cellcolor{yellow!60}{1.73}          & \cellcolor{yellow!60}{2.92}          & \cellcolor{yellow!60}{2.24}          & \cellcolor{yellow!60}{2.28} & \cellcolor{yellow!60}{1.60}          \\ \hline
\end{tabular}
}
\caption{The performance results on the Shadow dataset are presented. The dataset is segmented by images captured from different cameras, with "use all" representing the entire dataset. We report the mean and median angular errors for each subset. IFFCC employs the default settings, with a window size of 128 and an overlap size of 64. The model is trained using the blended illumination within each window as the ground truth (GT), and the corresponding error is computed.}
\label{shadowcamera}
\end{table*}

\begin{table*}[ht]
\centering
\resizebox{0.95\linewidth}{!}{
\setlength{\tabcolsep}{4.6pt}
\renewcommand{\arraystretch}{0.9}
\begin{tabular}{lcccccccccccc}
\hline
\multirow{2}{*}{Method}     & \multicolumn{4}{c}{Outdoor}                                   & \multicolumn{4}{c}{Indoor}                                    & \multicolumn{4}{c}{Nighttime}                                 \\ \cline{2-13} 
                            & mean          & median        & b25           & w25           & mean          & median        & b25           & w25           & mean          & median        & b25           & w25           \\ \hline
Huassin and Akbari\cite{hussain2018color}           & 13.19         & 12.89         & 7.10          & 19.12         & 14.84         & 13.22         & 8.43          & 20.18         & 16.35         & 14.96         & 9.75          & 22.33         \\
CRF(White-Patch)\cite{beigpour2013multi}             & 7.31          & 5.66          & 2.31          & 13.36         & 8.78          & 6.89          & 4.36          & 14.40         & 6.96          & 5.16          & 2.38          & 15.85         \\
Patch-based\cite{gijsenij2011color}       & 4.53          & 3.29          & 1.38          & 10.47         & 5.58          & 4.43          & 2.09          & 11.39         & 3.48          & 2.17          & 1.89          & 12.53         \\
Keypoint-based\cite{gijsenij2011color}  & 5.77          & 3.88          & 1.24          & 11.78         & 6.25          & 4.04          & 1.79          & 12.60         & 3.99          & 4.10          & 1.49          & 13.08         \\
Superpixel-based\cite{gijsenij2011color}             & 4.73          & 3.01          & 1.33          & 8.21          & 5.18          & 3.49          & 1.92          & 9.24          & 4.63          & 2.98          & 1.34          & 9.82          \\
IFFCC                       & \cellcolor{pink!60}{\textbf{1.74}} & \cellcolor{pink!60}{\textbf{1.28}} & \cellcolor{pink!60}{\textbf{0.42}} & \cellcolor{pink!60}{\textbf{3.83}} & \cellcolor{pink!60}{\textbf{3.04}} & \cellcolor{pink!60}{\textbf{2.14}} & \cellcolor{pink!60}{\textbf{0.66}} & \cellcolor{pink!60}{\textbf{6.71}} & \cellcolor{pink!60}{\textbf{2.85}} & \cellcolor{pink!60}{\textbf{1.53}} & \cellcolor{pink!60}{\textbf{0.54}} & \cellcolor{pink!60}{\textbf{7.43}} \\ \hline
Bianco et al.\cite{bianco2017single}               & 4.86          & 2.36          & 1.98          & 11.75         & 7.42          & 5.62          & 3.38          & 12.50         & 8.22          & 5.78          & 3.10          & 13.68         \\
HypNet/SelNet\cite{shi2016deep}               & 6.09          & 4.08          & 1.77          & 12.31         & 7.07          & 5.27          & 2.98          & 11.49         & 4.76          & 3.21          & 2.20          & 12.59         \\
Domislovic et al.\cite{domislovic2021outdoor}           & \cellcolor{yellow!60}{1.78}          & \cellcolor{yellow!60}{1.46}          & \cellcolor{yellow!60}{0.65}          & \cellcolor{yellow!60}{4.34}          & \cellcolor{yellow!60}{4.45}          & \cellcolor{yellow!60}{3.31}          & \cellcolor{yellow!60}{0.98}          & \cellcolor{yellow!60}{7.46}          & \cellcolor{yellow!60}{3.68}          & \cellcolor{yellow!60}{1.89}          & \cellcolor{yellow!60}{0.83}          & \cellcolor{yellow!60}{8.22}          \\ \hline
\end{tabular}
}
\caption{The performance results on the Shadow dataset are presented. The dataset is segmented by images captured in different scenes, with each scene including images taken by five different cameras. We report the mean and median angular errors for each subset. IFFCC uses the default settings, with a window size of 128 and an overlap size of 64. The model is trained using the blended illumination within each window as the ground truth (GT), and the corresponding error is computed.}
\label{type}
\end{table*}

\section{Experiment}
\subsection{Experimental Setup}
We evaluated our proposed IFFCC on two widely used multi-light source datasets: the LSMI dataset \cite{large} and the Shadow dataset \cite{domislovic2023shadows}. The LSMI dataset includes over 7,486 images captured by three different cameras in a variety of scenes, each containing multiple light sources. The Shadow dataset consists of 2,500 images taken in both indoor and outdoor settings, with each image accompanied by a binary segmentation mask, where each region is illuminated by a single light source. In all experiments, full-sized training images were randomly cropped to \(128\times128\), and testing was conducted on \(256\times256\) images. All images were black-level corrected, and calibration objects were masked.

IFFCC’s computational efficiency enables CPU-only training and inference, eliminating the need for a GPU. All experiments were conducted on an Intel Xeon Gold 6258R CPU. Performance was evaluated using standard error metrics, including mean and median angular errors and the arithmetic means of the first and third quartiles (“best 25\%” and “worst 25\%”). Additionally, we compared model parameter counts and CPU inference times for 256x256 images.
Since our method operates on patches, we calculate blended illumination values from the ground truth illumination map within a sliding window, using these as the ground truth for each patch. The impact of different training approaches is discussed in the ablation study.

During training, we input a set of enhanced histograms derived from filtered images that ensure non-negativity and intensity scaling consistency. Each image includes a local absolute deviation measure. Using the preconditioned frequency-domain optimization from FFCC \cite{barron2017fast}, training runs for 64 iterations with a histogram size of \(64 \times 64\).

\begin{figure*}[t]
    \centering
    \includegraphics[width=1\linewidth]{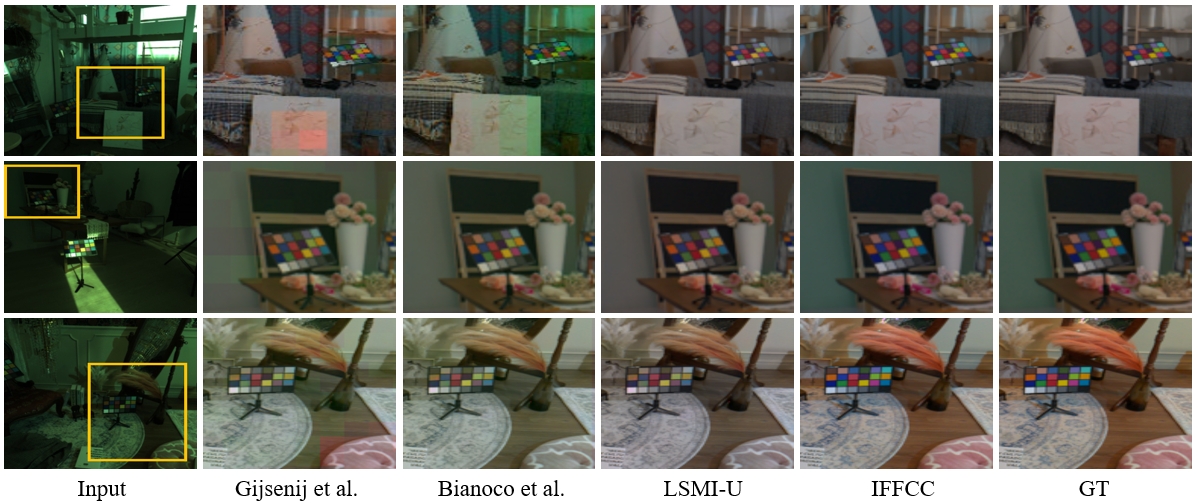}
    \caption{Visualization of the comparison results on the LSMI dataset. For IFFCC, the window size is set to 128 with an overlap of 64. In other patch-based methods, the window size is set to the default value of 32.}
    \label{f1}
\end{figure*}
\begin{figure*}[t]
    \centering
    \includegraphics[width=1\linewidth]{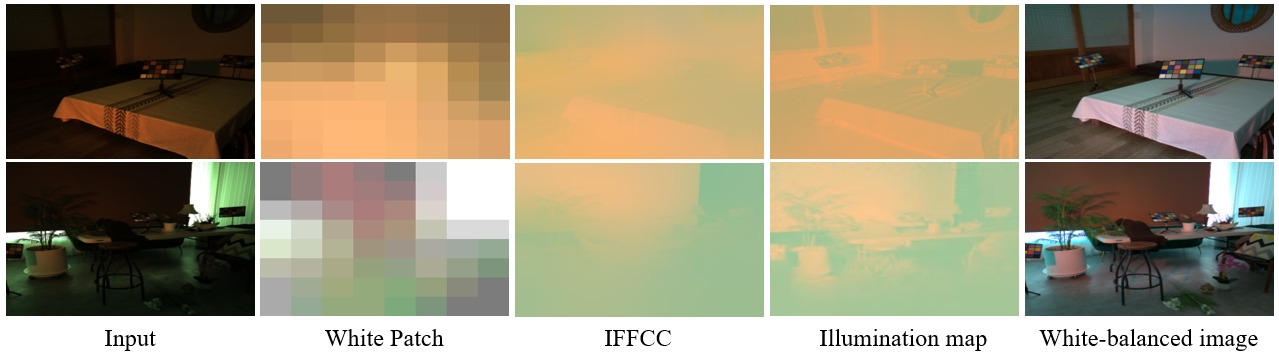}
    \caption{Visualization of illumination comparison for different patch-based methods. The window size is 32, with an overlap size of 16.}
    \label{f2}
\end{figure*}

\subsection{Results and Comparisons}
\textbf{Quantitative Comparison} We evaluated IFFCC against three traditional methods \cite{hussain2018color,beigpour2013multi,gijsenij2011color} and three learning-based methods \cite{bianco2017single,shi2016deep,domislovic2021outdoor} on the Shadow dataset. The experimental results are presented in Tables \ref{shadowcamera} and \ref{type}. Table \ref{shadowcamera} organizes the Shadow dataset into subsets based on images captured by different cameras, while Table \ref{type} categorizes the dataset by various scene types, each containing photos taken with five distinct cameras. Despite comparisons with advanced network-based methods, IFFCC achieves state-of-the-art performance. 

The results in Table \ref{type} highlight the strong generalization ability of IFFCC. Network-based methods generally perform worse than IFFCC on multi-camera datasets, where each camera's unique spectral sensitivities complicate learning a single mapping across different devices. IFFCC generalizes well across multiple cameras because it leverages an integral log-chroma histogram approach, which directly models chromaticity distributions without being overly reliant on the specific spectral sensitivities of individual cameras. This histogram-based method effectively captures illumination variations across different scenes and cameras, allowing IFFCC to produce robust predictions without requiring extensive fine-tuning for each camera’s unique spectral characteristics.

The experimental results on the LSMI dataset, presented in Tables \ref{multi}, and \ref{mixed}, provide a comparative analysis of patch-based and pixel-level methods. The test set is categorized into Multi (multiple light sources), and Mixed (comprising both single and multiple light sources). Among patch-based approaches, IFFCC demonstrates state-of-the-art performance, achieving results comparable to those of leading pixel-level methods, including AID \cite{kim2024attentive}. Notably, IFFCC’s compact parameter count and rapid processing speed render it highly suitable for deployment on mobile and resource-constrained devices.

We also compared the processing time for a 256×256 image in a multi-illuminant scene using FFCC and IFFCC, as shown in Table \ref{comffcc}. As the number of windows increases, IFFCC significantly outperforms FFCC in processing speed.

\noindent\textbf{Qualitative Comparison} 
Fig.\ref{f1} visually compares our method against alternative approaches on the LSMI dataset. To enhance visualization quality, color correction, gamma correction, and tone mapping were applied to the white-balanced images. The results indicate that IFFCC achieves more accurate white balance restoration than competing methods.
We also present a comparison of illumination maps between our method and patch-based approaches, as shown in Fig.\ref{f2}. The window size is set to \(32 \times 32\) with an overlap of 16. The results demonstrate that IFFCC achieves a more precise and smoother illumination prediction.

\begin{table}[t]
\resizebox{0.97\linewidth}{!}{
\setlength{\tabcolsep}{0.6pt}
\renewcommand{\arraystretch}{1.1}
\begin{tabular}{cccccccccc}
\hline
\multicolumn{2}{c}{\multirow{3}{*}{Method}}                                        & \multicolumn{6}{c}{Multi}                                                         & \multirow{3}{*}{\begin{tabular}[c]{@{}c@{}}Params\\ (M)\end{tabular}} & \multirow{3}{*}{\begin{tabular}[c]{@{}c@{}}Time\\ (s)\end{tabular}} \\ \cline{3-8}
\multicolumn{2}{c}{}                                                               & \multicolumn{2}{c}{Galaxy} & \multicolumn{2}{c}{Nikon} & \multicolumn{2}{c}{Sony} &                                                                       &                                                                     \\
\multicolumn{2}{c}{}                                                               & mean         & med.        & mean        & med.        & mean        & med.       &                                                                       &                                                                     \\ \hline
\multirow{3}{*}{\begin{tabular}[c]{@{}c@{}}Patch\\ -level\end{tabular}} & Gijsenij\cite{icc} & 12.38        & 9.57        & 9.31        & 6.59        & 11.19       & 8.14       & -                                                                     & 1.3                                                                 \\
                                                                        & Bianoco\cite{bianco2017single}  & 5.56         & 4.34        & 4.65        & 4.19        & 4.38        & 3.94       & \cellcolor{yellow!60}{0.16}                                                                  & -                                                                   \\
                                                                        & IFFCC    & \cellcolor{yellow!60}{2.48}         & \cellcolor{yellow!60}{1.90}        & \cellcolor{yellow!60}{2.30}        & \cellcolor{yellow!60}{1.50}        & \cellcolor{yellow!60}{2.48}        & \cellcolor{yellow!60}{1.97}       & \cellcolor{pink!60}{0.012}                                                                  & \cellcolor{pink!60}{0.03}                                                                \\ \hline
\multirow{4}{*}{\begin{tabular}[c]{@{}c@{}}Pixel\\ -level\end{tabular}} & Pix2Pix\cite{isola2017image}  & 4.28         & 2.63        & 4.18        & 2.76        & 4.37        & 3.26       & 5.4                                                                   & 0.48                                                                \\
                                                                        & LSMI-H\cite{large}   & 3.13         & 2.70        & 3.20        & 3.01        & 3.65        & 3.33       & 0.52                                                                  & \cellcolor{yellow!60}{0.28}                                                                \\
                                                                        & LSMI-U\cite{large}   & 2.85         & 2.55        & 2.36        & 1.84        & 2.76        & 2.38       & 5.7                                                                   & 0.51                                                                \\
                                                                        & AID\cite{kim2024attentive}      & \cellcolor{pink!60}{2.03}         & \cellcolor{pink!60}{1.43}        & \cellcolor{pink!60}{2.26}        & \cellcolor{pink!60}{1.39}        & \cellcolor{pink!60}{2.16}        & \cellcolor{pink!60}{1.64}       & 6.4                                                                   & \(>\)1                                                                  \\ \hline
\end{tabular}
}
\caption{The performance results on the multi-illumination images from the LSMI dataset are presented, with testing time corresponding to the processing time for 256×256 images on a CPU.}
\label{multi}
\end{table}

% \usepackage{multirow}
\begin{table}[t]
\resizebox{0.97\linewidth}{!}{
\setlength{\tabcolsep}{0.6pt}
\renewcommand{\arraystretch}{1.1}
\begin{tabular}{cccccccccc}
\hline
\multicolumn{2}{c}{\multirow{3}{*}{Method}}                                        & \multicolumn{6}{c}{Mixed}                                                         & \multirow{3}{*}{\begin{tabular}[c]{@{}c@{}}Params\\ (M)\end{tabular}} & \multirow{3}{*}{\begin{tabular}[c]{@{}c@{}}Time\\ (s)\end{tabular}} \\ \cline{3-8}
\multicolumn{2}{c}{}                                                               & \multicolumn{2}{c}{Galaxy} & \multicolumn{2}{c}{Nikon} & \multicolumn{2}{c}{Sony} &                                                                       &                                                                     \\
\multicolumn{2}{c}{}                                                               & mean         & med.        & mean        & med.        & mean        & med.       &                                                                       &                                                                     \\ \hline
\multirow{3}{*}{\begin{tabular}[c]{@{}c@{}}Patch\\ -level\end{tabular}} & Gijsenij\cite{icc} & 10.09        & 7.43        & 9.21        & 5.83        & 9.46        & 7.13       & -                                                                     & 1.3                                                                 \\
                                                                        & Bianoco\cite{bianco2017single}  & 4.89         & 3.83        & 3.93        & 3.48        & 3.84        & 3.21       & \cellcolor{yellow!60}{0.16}                                                                  & -                                                                   \\
                                                                        & IFFCC    & \cellcolor{yellow!60}{1.98}         & \cellcolor{yellow!60}{1.79}        & \cellcolor{yellow!60}{2.11}        & \cellcolor{yellow!60}{1.48}        & \cellcolor{yellow!60}{1.93}        & \cellcolor{yellow!60}{1.72}       & \cellcolor{pink!60}{0.012}                                                                 & \cellcolor{pink!60}{0.03}                                                                \\ \hline
\multirow{4}{*}{\begin{tabular}[c]{@{}c@{}}Pixel\\ -level\end{tabular}} & Pix2Pix\cite{isola2017image}  & 5.66         & 2.44        & 5.41        & 2.49        & 4.20        & 2.20       & 5.4                                                                   & 0.48                                                                \\
                                                                        & LSMI-H\cite{large}   & 3.06         & 2.54        & 2.92        & 2.63        & 3.21        & 2.85       & 0.52                                                                  & \cellcolor{yellow!60}{0.28}                                                                \\
                                                                        & LSMI-U\cite{large}   & 2.63         & 1.91        & 2.16        & 1.55        & 2.68        & 2.29       & 5.7                                                                   & 0.51                                                                \\
                                                                        & AID\cite{kim2024attentive}      & \cellcolor{pink!60}{1.63}         & \cellcolor{pink!60}{1.32}        & \cellcolor{pink!60}{1.76}        & \cellcolor{pink!60}{1.14}        & \cellcolor{pink!60}{1.68}        & \cellcolor{pink!60}{1.44}       & 6.4                                                                   & \(>\)1                                                                  \\ \hline
\end{tabular}
}
\caption{The performance results on the mixed-illumination images (containing both single and multiple light sources) from the LSMI dataset are presented, with testing time corresponding to the processing time for 256×256 images on a CPU.}
\label{mixed}
\end{table}

\begin{table}[t]
\resizebox{1\linewidth}{!}{
\begin{tabular}{ccccccc}
\hline
      & {[}32,0{]} & {[}32,16{]} & {[}64,32{]} & {[}64,48{]} & {[}128,64{]} & {[}128,96{]} \\ \hline
FFCC  & 162ms      & 450ms       & 203ms       & 689ms       & 88ms         & 234ms        \\
IFFCC & \textbf{34ms}       & \textbf{46ms}        & \textbf{29ms}        & \textbf{40ms}        & \textbf{27ms}         & \textbf{33ms}         \\ \hline
\end{tabular}
}
\caption{Comparison of average testing time for FFCC and IFFCC on the Shadow dataset, evaluated across different window and overlap sizes. FFCC employs a conventional histogram computation approach.}
\label{comffcc}
\end{table}
\vspace{-1mm}

\subsection{Ablation Study}
\begin{figure}
    \centering
    \includegraphics[width=1\linewidth]{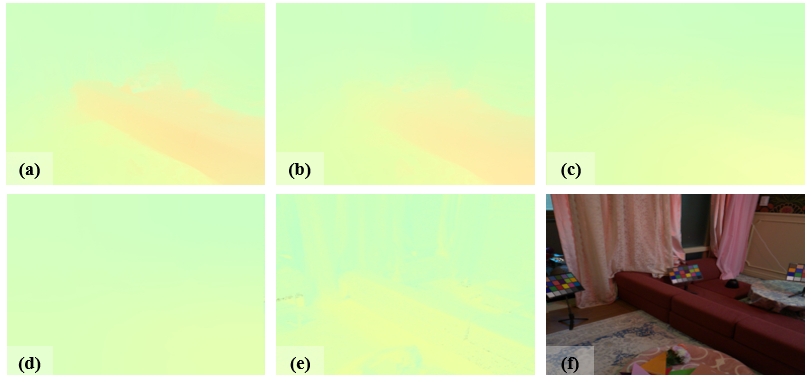}
    \caption{(a)-(d) show illumination maps generated with different window sizes ([32, 16], [64, 32], [128, 64], [192, 128]), each processed with smooth transition adjustments. (e) presents the ground truth (GT) illumination map, and (f) displays the white-balanced image obtained by combining the input image with the GT.}
    \label{ablation}
\end{figure}

Table \ref{t6} presents results from ablation experiments analyzing the impact of various training strategies, window sizes, and overlap regions. Here, “B” represents training with a blend of illumination maps as the ground truth (GT), and “M” uses the dominant light source as GT. Parameters “ws” and “os” indicate window and overlap sizes, respectively. Findings reveal that smaller window sizes, with limited information, reduce prediction accuracy. However, the blended illumination strategy provides improved accuracy compared to using a single dominant light source. Visual comparisons (Fig. \ref{ablation}) show that smaller windows enhance texture details in the predicted illumination maps but with lower accuracy. Conversely, larger windows (e.g., [192, 128]) reduce error metrics by capturing more information, yet they may blur boundaries in multi-illuminant regions, compromising edge clarity.

\begin{table}[t]
\centering
\resizebox{0.8\linewidth}{!}{
\setlength{\tabcolsep}{7.5pt}
\renewcommand{\arraystretch}{0.9}
\begin{tabular}{cccccc}
\hline
type & {[}ws, os{]}   & mean & median & b25  & w25  \\ \hline
B    & {[}32, 16{]}   & 2.65 & 2.17 & 1.37 & 4.65 \\
B    & {[}64, 0{]}    & 2.35 & 2.01 & 1.13 & 4.31 \\
M    & {[}64, 32{]}   & 2.53 & 2.04 & 1.25 & 4.83   \\
B    & {[}64, 32{]}   & 2.37 & 1.87 & 1.16 & 4.38 \\
M    & {[}128, 0{]}   & 2.46 & 2.03 & 1.04 & 4.78   \\
B    & {[}128, 0{]}   & 2.16 & 1.68 & 0.91 & 4.23 \\
M    & {[}128, 32{]}  & 2.50 & 1.88 & 1.04 & 4.91  \\
B    & {[}128, 32{]}  & 2.14 & 1.64 & 0.85 & 4.30 \\
M    & {[}128, 64{]}  & 2.42 & 1.95 & 0.93 & 4.71     \\
B    & {[}128, 64{]}  & 2.06 & 1.54 & 0.81 & 4.19 \\
B    & {[}128, 96{]}  & 2.13 & 1.59 & 0.83 & 4.25 \\
B    & {[}192, 128{]} & 1.98 & 1.34 & 0.72 & 4.08  \\ \hline
\end{tabular}
}
\caption{Ablation study on the Shadow dataset (Canon5d), reporting the impact of different window sizes, overlap sizes, and training strategies on accuracy.}
\label{t6}
\end{table}

\section{Conclusion}

We propose IFFCC, a practical solution for achieving accurate, real-time auto white balance (AWB) in multi-illuminant environments. Built upon the Fast Fourier Color Constancy (FFCC) algorithm, IFFCC leverages an integral UV histogram and parallelized Fourier-based convolution to rapidly estimate illumination across different image regions, addressing the need for localized, adaptive AWB. Results on multiple datasets demonstrate that IFFCC delivers accuracy comparable to or surpassing pixel-level neural network methods, while reducing the parameter count by over 400 times and speeding up processing by a factor of 20–100. These advantages make IFFCC highly suitable for resource-limited, real-time applications such as video streaming.

We believe that FFCC-based methods, including IFFCC, will continue to enrich the field and attract increased attention across a broader range of ISP applications.
{

    \small
    \bibliographystyle{ieeenat_fullname}
    % \bibliography{main}
}
% \bibliographystyle{ieeenat_fullname}
% \bibliography{main}

\end{document}